\documentclass[12pt,preprint]{aastex}
\begin{document}

\title{Convection in galaxy cluster plasmas driven by active galactic nuclei and
cosmic ray buoyancy}
\author{Benjamin D. G. Chandran}
\email{benjamin-chandran@uiowa.edu} 
\affil{Department of Physics \& Astronomy, University of Iowa}

\begin{abstract}

Turbulent heating may play an important role in
galaxy-cluster plasmas, but if turbulent heating is to
balance radiative cooling in a quasi-steady state, some
mechanism must regulate the turbulent velocity so that it is
neither too large nor too small. This paper explores one
possible regulating mechanism associated with an active
galactic nucleus at cluster center. A steady-state model for
the intracluster medium is presented in which radiative
cooling is balanced by a combination of turbulent heating
and thermal conduction. The turbulence is generated by
convection driven by the buoyancy of cosmic rays produced by
a central radio source. The cosmic-ray luminosity is powered
by the accretion of intracluster plasma onto a central black
hole.  The model makes the rather extreme assumption that
the cosmic rays and thermal plasma are completely
mixed. Although the intracluster medium is convectively
unstable near cluster center in the model solutions, the
specific entropy of the thermal plasma still increases
outwards because of the cosmic-ray modification to the
stability criterion.  The model provides a self-consistent
calculation of the turbulent velocity as a function of
position, but fails to reproduce the steep central density
profiles observed in clusters. The principal difficulty is
that in order for the fully mixed intracluster medium to
become convectively unstable, the cosmic-ray pressure must
become comparable to or greater than the thermal pressure
within the convective region. The large cosmic-ray pressure
gradient then provides much of the support against gravity,
reducing the thermal pressure gradient near cluster center
and decreasing the central plasma density gradient. A more
realistic AGN-feedback model of intracluster turbulence in
which relativistic and thermal plasmas are only partially
mixed may have greater success.

\end{abstract}
\maketitle

\section{Introduction}

Recent x-ray observations show that some source of heating inhibits
cooling of intracluster plasma to temperatures below roughly one-third
of the average cluster temperature (Peterson et~al 2001, Peterson
et~al 2003, B\"{o}hringer et~al 2001, Tamura et~al
2001, Molendi \& Pizzolato 2001). One of the relevant heating
mechanisms is conduction of heat from a cluster's hot outer region
towards cluster center. Fabian et~al (2002), Voigt et~al (2002), and
Zakamska \& Narayan (2003) showed that thermal conduction can balance
radiative cooling in many clusters if the thermal
conductivity~$\kappa_T$ is in the range $0.1-1.0\kappa_{\rm S}$,
where~$\kappa_{\rm S}$ is the Spitzer thermal conductivity.  However,
models in which cooling is balanced entirely by conduction face some
difficulties. Different clusters need different values of~$\kappa_T$
to fit the observations, and it is not clear why $\kappa_T$ would vary
in the required way. Relatively cool cluster plasmas can require
values of~$\kappa_T$ in excess of~$\kappa_{\rm S}$ for conduction to
balance cooling, values which are difficult if not impossible to
justify.  Theoretical studies of heat transport in tangled
intracluster magnetic fields find values of~$\kappa_T$ towards the
lower end of the range of required values (Narayan \& Medvedev 2001,
Chandran \& Maron 2004, Bessho,
Maron, \& Chandran 2004, and references therein).  In addition, pure
conduction models are thermally unstable, although the implications
for clusters are debated since the instability growth time is long
($\sim 2-5$~Gyr for $\kappa_{\rm T} \sim 0.2-0.4 \kappa_{\rm S}$) (Kim
\& Narayan 2003, Soker 2003).

Other heating mechanisms include dissipation of turbulent motions
(Loewenstein \& Fabian 1990, Churazov et~al 2004, Dennis \& Chandran
2004) and turbulent mixing (Cho et~al 2003, Voigt \& Fabian 2004, Kim
\& Narayan 2003, Dennis \& Chandran 2004, Narayan \& Kim
2004). Observational evidence of plasma motion in the Perseus cluster
at roughly half the sound speed~$c_{\rm s}$ (Churazov et~al~2004)
provides support for models invoking turbulent heating, but also poses
a theoretical challenge for the following reason. As the rms turbulent
velocity~$u$ varies from~$0.1 c_{\rm s}$ to~$c_{\rm s}$, and as the
dominant velocity length scale at radius~$r$ varies from~$0.1r$
to~$r$, the turbulent heating rate in a typical cluster varies from a
value much smaller than the radiative cooling rate~$R$ to a value much
larger than~$R$ (Dennis \& Chandran 2004). If turbulent heating in
fact balances radiative cooling, an explanation is needed for how the
turbulence amplitude is fine-tuned to the required value.

A number of authors have considered plasma heating by an active
galactic nucleus (AGN) at a cluster's center.  Several mechanisms have
been considered for transferring AGN power to the ambient plasma,
including buoyantly rising bubbles of cosmic rays or centrally heated
gas (Tabor \& Binney 1993, Churazov et~al~2000, Churazov et~al~2001,
Begelman~2001a, 2001b, Reynolds~2001, Bruggen et~al~2002,
Begelman~2002, Ruszkowski \& Begelman~2002, Fabian et~al~2003, Mathews
et~al~2003, Reynolds et~al~2004), wave-mediated plasma heating by cosmic-ray electrons
(Rosner \& Tucker~1989) or cosmic-ray protons (B\"{o}hringer \&
Morfill~1988, Loewenstein, Zweibel, \& Begelman~1991), turbulence
(Loewenstein \& Fabian~1990, Churazov et~al~2004), Compton heating
(Binney \& Tabor 1995, Ciotti \& Ostriker~1997,~2001), and jet
mechanical luminosity and shocks~(Binney \& Tabor~1995). An attractive
feature of AGN-heating models is that the heating rate increases with
the mass accretion rate of intracluster plasma, which allows the
models to produce globally stable equilibria.  If the mass accretion
rate rises above the equilibrium value, AGN heating increases, and
radiative cooling of the ICM is less able to cause the net cooling
that drives mass accretion onto the central supermassive black
hole. If the mass accretion rate falls below the equilibrium value,
the net cooling rate rises, restoring the mass accretion rate to its
equilibrium level.  If the thermal conductivity is not dramatically
reduced relative to the Spitzer value, small-scale thermal
instabilities are also suppressed (Rosner \& Tucker 1989, Ruszkowski
\& Begelman 2002, Zakamska \& Narayan 2003). A powerful motivation for
AGN-heating models is the observation that almost all clusters with
strongly cooling cores possess active central radio sources (Eilek
2003).

The present paper builds upon previous studies of turbulent heating
and AGN feedback in clusters, and explores one possible explanation
for how the turbulent velocity could achieve the value needed to
balance radiative cooling.  A steady-state model of the ICM is
presented in which radiative cooling causes inflow of intracluster
plasma and accretion onto a central supermassive black hole, leading
to cosmic-ray production by a central radio source. The cosmic-ray
luminosity~$L_{\rm cr}$ is taken to be proportional to the mass accretion
rate~$\dot{M}$, and it is assumed that the cosmic-rays mix completely
into the thermal intracluster plasma.\footnote{See Ensslin (2003)
for a detailed discussion of the escape of cosmic rays from radio-galaxy
cocoons.} Since the cosmic rays provide
pressure without noticeably increasing the density of the ICM, they
increase the buoyancy of the plasma, eventually leading to
convection. Convection heats the plasma in three ways, through the
viscous dissipation of turbulent motions, by mixing hot plasma from
the outer regions of a cluster in towards cluster center, and by
providing a vehicle for cosmic-ray pressure to do work on the thermal
plasma.  Convection is treated with a two-fluid (thermal plasma and
cosmic ray) mixing length theory developed in section~\ref{sec:mod}.
The turbulent velocity and turbulent heating rate increase with
increasing~$\dot{M}$ and~$L_{\rm cr}$. In equilibrium, $\dot{M}$
and~$L_{\rm cr}$ attain those values for which turbulent heating and
thermal conduction balance radiative cooling.  This equilibrium is
expected to be stable for the same reasons as other
AGN-feedback/thermal-conduction models.

Once the outer temperature and density are specified, the model can be
used to calculate the density, temperature, and cosmic-ray pressure
throughout the cluster. The model density profiles, however, are not as
steep as observed profiles within the central~$\sim 50$~kpc,
where the ICM is convective. The reason for this discrepancy is that
the cosmic-ray pressure has to be comparable to or greater than the
thermal pressure within the convective region in order for cosmic-ray
buoyancy to make the ICM convectively unstable.  The large cosmic-ray
pressure gradient provides much of the support of the ICM against
gravity, reducing the thermal pressure gradient, and decreasing the
central plasma density. A similar discrepancy was found in an
MHD-wave-mediated cosmic-ray heating model (Loewenstein et~al~1991),
which also involved significant nonthermal pressure.  The present
model also provides a self-consistent calculation of the profiles of
the turbulent velocity and turbulent heating rates in the~ICM. The
resulting velocities are subsonic, implying that convective regions
are fairly close to marginal convective stability. However, because
the cosmic-ray pressure modifies the stability criterion, the specific
entropy of thermal plasma still increases outwards.

The discrepancy between the model and observations
may be linked to the assumption that cosmic
rays are completely mixed into the thermal plasma.  This assumption,
which is made to simplify the analysis, is inconsistent with the
x-ray cavities (depressions in x-ray emission often associated with
enhanced synchrotron emission) observed in roughly one-fourth of the
clusters in the Chandra archive (Birzan et~al~2004). When mixing is
incomplete, cosmic-ray pressure is less able to support the thermal
plasma, the thermal pressure gradient has to be larger, and therefore
the central plasma density has to be larger. A model that accounts for
incomplete mixing may have greater success.

The remainder of the paper is organized as follows. In
section~\ref{sec:mod}, the model is described in detail and 
example solutions are compared to observations of Abell~478.
Section~\ref{sec:dis} summarizes the results of the paper.

\section{The two-fluid convection model}
\label{sec:mod} 

In the cooling flow model of intracluster plasmas (Fabian 1994), entropy-generating
heat sources are neglected, and radiative losses are balanced by
inward advection of plasma internal energy as well as gravitational and
$pdV$~work.  The resulting average radial velocity corresponds to mass
accretion rates above~$1000 M_\sun \mbox{ yr}^{-1}$ in some
clusters. In contrast, in the present model, radiative cooling is
balanced by heating from turbulence and thermal conduction, leading to
much smaller radial velocities and mass accretion rates.
The average fluid velocity is neglected in the fluid
equations, and the mass accretion rate is set equal to Bondi (1952)
rate corresponding to the equilibrium plasma parameters at cluster
center,
\begin{equation}
\dot{M} = \frac{\pi G^2 M_{\rm bh}^2 \rho(0)}{c_{\rm s,ad}(0) ^3},
\end{equation} 
where $M_{\rm bh}$ is the mass of a supermassive black hole at cluster
center, $\rho(0)$ is the central density, and $c_{\rm s,ad}(0)$ is the
adiabatic sound speed at cluster center.\footnote{See Quataert \&
Narayan (2000), Nulsen (2003), and B\"{o}hringer et~al (2003) for
further discussions of Bondi accretion in clusters.}

The inferred mass accretion rate is used to determine the
cosmic-ray luminosity through the equation
\begin{equation}
L_{\rm cr} = \eta \dot{M} c^2,
\label{eq:lcr} 
\end{equation} 
where $\eta$ is a dimensionless efficiency factor.  The spatial
distribution of cosmic-ray injection into the ICM is a major
uncertainty in the model. Some clues are provided by radio
observations, which show that cluster-center radio sources (CCRS)
differ morphologically from radio sources in other environments. As
discussed by Eilek (2003), roughly half of the CCRS in a sample of 250
sources studied by Owen \& Ledlow (1997) are ``amorphous,'' or
quasi-isotropic, presumably due to jet disruption by the comparatively
high-pressure cluster-core plasma. With the exception of Hydra~A, the
CCRS in the Owen-Ledlow (1997) study are smaller than
non-cluster-center sources, with most extending less than 50~kpc from
cluster center (Eilek 2003).  Complicating matters is the possibility
that cosmic-ray bubbles rise buoyantly away from their acceleration
site before mixing into the thermal plasma, effectively distributing
cosmic-ray injection over a larger volume.  In this paper, the
cosmic-ray energy introduced into the intracluster medium per unit
volume per unit time is taken to be 
\begin{equation}
S(r) = S_0 e^{-r^2/r_s^2},
\label{eq:defS} 
\end{equation} 
where the constant~$r_s$ is a free parameter and
the constant~$S_0$ is determined from the equation
$L_{\rm cr} = 4\pi \int_0^\infty dr \, r^2 S(r)$ and
equation~(\ref{eq:lcr}).

The magnetic pressure is taken to be much less than the thermal
pressure. Magnetic terms are then neglected in the fluid equations,
and the effects of the magnetic field on convection are
ignored.\footnote{See Pen et~al~(2003) for a very different picture
of the interplay between magnetic fields and convection in 
quasi-spherical accretion, in which convective
motions are resisted by magnetic tension.}
Although the magnetic field does not affect the bulk fluid
velocity, for any plausible field strength the field plays an important
role in heat transport by constraining charged particles to move
primarily along field lines. Since the magnetic field in clusters is
disordered and probably turbulent (Kronberg 1994, Taylor et~al~2001,
2002), chaotic field-line trajectories inhibit large radial particle
excursions, suppressing radial heat transport to some degree.  
In this paper, the thermal conductivity is set equal to the Spitzer~(1962)
value for a non-magnetized plasma,~$\kappa_{\rm S}$,
multiplied by a suppression factor~$\theta$,
\begin{equation}
\kappa_T = \theta \kappa_{\rm S},
\end{equation} 
where
\begin{equation}
\frac{\kappa_{\rm S}}{n_e k_B} = 9.2 \times 10^{30} \left(
\frac{k_B T}{5 \mbox{ keV}}\right)^{5/2}
\left(\frac{10^{-2} \mbox{ cm}^{-3}}{n_e}\right)
\left(\frac{37}{\ln \Lambda_{\rm c}}\right) \frac{\mbox{ cm}^{2}}{\mbox{ s}},
\label{eq:kappas} 
\end{equation} 
$n_e$ is the electron density,~$T$ is the temperature, $k_B$ is the
Boltzmann constant, and $\ln \Lambda_{\rm c}$ is the Coulomb
logarithm. Recent theoretical studies suggest that~$\theta\sim 0.2$
(Narayan \& Medvedev 2001, Chandran \& Maron 2004, Maron, Chandran, \& Blackman~2004).
The heating rate per unit volume from thermal conduction is 
\begin{equation}
H_{\rm tc} = \nabla \cdot (\kappa_T \nabla T).
\label{eq:defHtc} 
\end{equation} 
Following Tozzi \& Norman (2001) and Ruszkowski \& Begelman (2002), the
radiative cooling per unit volume per unit time
for free-free and line emission is taken to be
\begin{equation}
R = n_i n_e \left[0.0086\left(\frac{k_B T}{\mbox{1 keV}}\right)^{-1.7}
+ 0.058 \left(\frac{k_B T}{\mbox{1 keV}}\right)^{0.5} + 0.063\right]
\cdot 10^{-22} \mbox{ ergs}\mbox{ cm}^3 \mbox{ s}^{-1},
\label{eq:R1}
\end{equation} 
where $n_i$ is the ion density
and the numerical constants correspond to 30\% solar metallicity.
Radiative cooling of cosmic rays is ignored, which is reasonable if
protons make the dominant contribution to the cosmic-ray pressure.
Cosmic rays are assumed to diffuse relative to the thermal plasma, but
the value of the cosmic-ray diffusion coefficient~$D_{\rm cr}$ in clusters is not known.
In the model, 
\begin{equation}
D_{\rm cr} = \sqrt{D_0^2 + v_d^2 r^2},
\label{eq:defDcr} 
\end{equation} 
where~$D_0$ and~$v_d$ are constants.

The thermal plasma and cosmic rays are treated as co-extensive fluids
that interact only in limited ways. Wave pitch-angle scattering causes
the two fluids to move with the same bulk fluid velocity~${\bf v}$,
although the finite efficiency of scattering allows for cosmic-ray
diffusion as already discussed. Collisional and microphysical
collisionless energy exchange between the two fluids is neglected, but
the cosmic rays can heat the plasma, and vice versa, through~$p\,dV$
work. The thermal plasma is treated as a fluid of adiabatic index~$\gamma$,
\begin{equation}
p = (\gamma -1) \rho \epsilon,
\label{eq:est} 
\end{equation} 
where~$p$ is the plasma pressure, $\rho$ is the plasma density,
and~$\epsilon$ is the plasma internal energy per unit mass.
The cosmic rays are treated as a fluid of adiabatic index~$\gamma_{\rm cr}$,
\begin{equation}
p_{\rm cr} = (\gamma_{\rm cr} -1) \rho_{\rm cr} \epsilon_{\rm cr},
\label{eq:escr} 
\end{equation} 
where~$p_{\rm cr}$ is the cosmic-ray pressure, $\rho_{\rm cr}$ is the
cosmic-ray density, and~$\epsilon_{\rm cr}$ is the cosmic-ray internal
energy per unit mass.  Both~$\gamma$ and~$\gamma_{\rm cr}$ are treated
as constants (although the adiabatic index of a fluid should really
vary with temperature from~5/3 in the non-relativistic limit to~4/3 in
the ultra-relativistic limit). 

The first law of thermodynamics can be written
\begin{equation}
\frac{d\epsilon}{dt} = H + \frac{p}{\rho^2} \frac{d\rho}{dt},
\label{eq:flt} 
\end{equation} 
where~$d/dt \equiv (\partial/\partial t \,+ \,{\bf v} \cdot \nabla)$,
and~$H$ is the net heating per unit mass per unit time.
Equations~(\ref{eq:est}) and (\ref{eq:flt}) together give
\begin{equation}
\frac{1}{\gamma-1} \left(
\frac{dp}{dt} - \frac{\gamma p}{\rho} \frac{d\rho}{dt}\right)
= \rho H.
\label{eq:eqni} 
\end{equation} 
The equation of continuity is
\begin{equation}
\frac{\partial \rho}{\partial t} + \nabla \cdot (\rho {\bf v}) = S_M,
\label{eq:cont1} 
\end{equation} 
where~$S_M$ is the rate at which mass is introduced (by, e.g., stellar
winds) per unit time per unit volume. Equations~(\ref{eq:eqni}) and (\ref{eq:cont1}) 
give
\begin{equation}
\frac{1}{\gamma -1} \left[\frac{\partial p}{\partial t}
+ \nabla \cdot ({\bf v} p) \right] = - p\nabla \cdot {\bf v}
+ \rho H + \frac{\gamma p S_M}{(\gamma-1) \rho}.
\label{eq:ene1} 
\end{equation} 
Similarly, 
\begin{equation}
\frac{1}{\gamma_{\rm cr} -1} \left[\frac{\partial p_{\rm cr}}{\partial
t} + \nabla \cdot ({\bf v} p_{\rm cr}) \right] = - p_{\rm cr}\nabla
\cdot {\bf v} + \rho_{\rm cr} H_{\rm cr} + \frac{\gamma_{\rm cr} p_{\rm cr}
S_{M, \rm cr}}{(\gamma_{\rm cr}-1) \rho_{\rm cr}},
\label{eq:enecr1} 
\end{equation} 
where $H_{\rm cr}$ and $S_{M, \rm cr}$ are, respectively, the cosmic-ray heating and
mass-injection rates. For the thermal plasma,
\begin{equation}
\rho H = H_{\rm diss} + H_{\rm tc} - R,
\label{eq:defH} 
\end{equation} 
where $H_{\rm diss}$ 
is the rate of heating from viscous dissipation 
of turbulent motions, quantified below in
equation~(\ref{eq:defHd}).  For simplicity,
it is assumed that
\begin{equation}
S_M = 0.
\label{eq:defSm} 
\end{equation} 
The right-hand side of equation~(\ref{eq:enecr1}) is simplified
by assuming that 
\begin{equation}
\rho_{\rm cr} H_{\rm cr} + \frac{\gamma_{\rm cr} p_{\rm cr}
S_{M, \rm cr}}{(\gamma_{\rm cr}-1) \rho_{\rm cr}}
= \nabla \cdot (D_{\rm cr} \nabla p_{\rm cr} ) + S(r),
\label{eq:defHcr} 
\end{equation} 
with $D_{\rm cr}$ given by equation~(\ref{eq:defDcr}) 
and $S(r)$ given by equation~(\ref{eq:defS}). The first
term on the right-hand side of equation~(\ref{eq:defHcr}) 
is a simplified representation of diffusive energy transport,
and the second term models the energy injection from
the central radio source. 

Convection and convective stability are treated using the following
simple two-fluid mixing length theory. Consider a parcel of plasma and cosmic
rays initially at a distance~$r=r_0$ from cluster center. The parcel
is displaced radially by an amount~$l$ (positive or negative)
to $r=r_1 = r_0 + l$, where~$|l|$ is the mixing length.
For simplicity, thermal conduction and cosmic-ray diffusion into and
out of the parcel are neglected at this stage, so that the parcel expands
adiabatically.  The parcel is then mixed with its
surroundings.  The values of~$\rho$, $p$, $\rho_{\rm cr}$, and~$p_{\rm cr}$
in the parcel are  initially the same as the average values at~$r=r_0$,
denoted~$\rho_0$, $p_0$, $\rho_{\rm cr,0}$ and~$p_{\rm cr, 0}$.
After the parcel is displaced radially
outwards, the new fluid quantities within the parcel are denoted
$\rho^\prime$, $p^\prime$, $\rho_{\rm cr}^\prime$, and~$p_{\rm
cr}^\prime$. The average fluid quantities at~$r=r_1$ are
denoted~$\rho_1$, $p_1$, $\rho_{\rm cr, 1}$, and~$p_{\rm cr, 1}$.  The
difference between the thermal-plasma density in the displaced parcel
and its surroundings at~$r=r_1$ is denoted
\begin{equation}
\Delta \rho = \rho^\prime - \rho_1.
\label{eq:defdrho} 
\end{equation} 
Similarly, $\Delta p = p^\prime - p_1$, etc.
Since the parcel expands adiabatically, 
\begin{equation}
p^\prime = p_0 \left(\frac{\rho^\prime}{\rho_0}\right)^\gamma,
\label{eq:pp1} 
\end{equation}
and
\begin{equation}
p_{\rm cr}^\prime = p_{\rm cr, 0}\left(\frac{\rho_{\rm cr}^\prime}{\rho_{\rm cr, 0}}\right)^{\gamma_{\rm cr}}.
\label{eq:pcrp1} 
\end{equation} 
The volume occupied by the thermal plasma and the volume occupied by
the cosmic rays expand by the same amount, which implies that
\begin{equation}
\frac{\rho^\prime}{\rho_0} = \frac{\rho_{\rm cr}^\prime}{\rho_{\rm cr, 0}}
\equiv 1 + \delta.
\label{eq:volrat} 
\end{equation} 
It is assumed that the turbulent velocities are subsonic, so that the
total pressure 
\begin{equation}
p_{\rm tot} = p + p_{\rm cr}
\label{eq:defptot} 
\end{equation} 
inside the parcel 
remains approximately the same as the total pressure outside the parcel. 
It is now assumed that~$|l| \ll r$, which implies that~$\delta \ll 1$.
To lowest order in~$|l|/r$,
equations~(\ref{eq:pp1}) through (\ref{eq:defptot}) give
\begin{equation} 
\delta = \frac{l}{\gamma p + \gamma_{\rm cr} p_{\rm cr}} \frac{dp_{\rm tot}}{dr},
\label{eq:valdelta} 
\end{equation} 
\begin{equation}
\Delta \rho = \frac{l\rho}{\gamma p + \gamma_{\rm cr} p_{\rm cr}} \frac{dp_{\rm tot}}
{dr} - l \frac{d\rho}{dr},
\label{eq:valdrho} 
\end{equation} 
\begin{equation}
\Delta p = \frac{l\gamma p}{\gamma p + \gamma_{\rm cr} p_{\rm cr}} 
\frac{d p_{\rm tot}}{dr} - l\frac{dp}{dr},
\label{eq:valdp} 
\end{equation} 
and
\begin{equation}
\Delta p_{\rm cr} = - \Delta p.
\label{eq:valdpcr} 
\end{equation} 
Since only the lowest-order terms have been kept, it is not necessary
to specify in equations~(\ref{eq:valdelta}) through (\ref{eq:valdp}) whether
$\rho$,~$p$, etc are evaluated at~$r_0$ or~$r_1$, and so 
the subscripts on $\rho$,~$p$, etc have been dropped.

Since the displacement~$l$ is a signed quantity, the
criterion for convective stability is 
\begin{equation}
\frac{\Delta \rho}{l} > 0.
\label{eq:stabcrit} 
\end{equation} 
If $p\gg p_{\rm cr}$, equation~(\ref{eq:stabcrit}) is equivalent
to the Schwarzchild criterion
\begin{equation}
\frac{d}{dr} \ln \left(\frac{p}{\rho^\gamma}\right) > 0.
\label{eq:stabcrit1} 
\end{equation} 
More generally, the ICM can be convectively unstable even when
the specific entropy of the thermal plasma increases outwards, provided
that $dp_{\rm cr}/dr$ is negative and~$|dp_{\rm cr}/dr|$ is sufficiently
large.

The rms turbulent velocity~$u$ 
is assumed to satisfy
\begin{equation}
\rho u^2 \simeq \left\{ \begin{array}{ll}
\displaystyle \frac{\Delta \rho \, gl}{8} & \mbox{ \hspace{0.3cm} if 
\hspace{0.3cm} $\displaystyle 
l^{-1}\Delta \rho  < 0$} \\
0 & \mbox{ \hspace{0.3cm} if \hspace{0.3cm} $\displaystyle l^{-1} \Delta \rho > 0$,}
\end{array}\right.
\label{eq:u1} 
\end{equation} 
where~$g$ is the gravitational acceleration. That is, the
bulk-flow kinetic energy of a moving parcel in a convective
region is approximately the buoyancy force on the fully
displaced parcel times the mixing length times the
standard numerical coefficient used in mixing-length theory
(Cox \& Giugli 1968).  Equation~(\ref{eq:u1}) is then
modified slightly so that $du/dr$ varies continuously to
zero as~$\Delta \rho$ increases through~zero:
\begin{equation}
 u^2 = \left\{ \begin{array}{ll}
\displaystyle u_0^2 + \frac{ q g l}{8}
 & \mbox{ \hspace{0.3cm} if \hspace{0.3cm} $
l^{-1} \Delta \rho  < 0$} \\
\displaystyle 
u_0^2 e^{-\sigma q} & \mbox{ \hspace{0.3cm} if \hspace{0.3cm} 
$ l^{-1}\Delta \rho  > 0$,}
\end{array}\right.
\label{eq:u2} 
\end{equation} 
where 
\begin{equation}
q = \frac{\Delta \rho}{\rho},
\label{eq:defq} 
\end{equation} 
\begin{equation}
\sigma =  - \frac{g l}{8 u_0^2},
\label{eq:defsigma} 
\end{equation} 
\begin{equation}
u_0 = u_0^\prime r,
\end{equation} 
and~$u_0^\prime$ is a constant chosen so that~$u_0$ remains much
smaller than the sound speed throughout the cluster.  The value
of~$\sigma$ in equation~(\ref{eq:defsigma}) is chosen so that
$\partial u^2/\partial q$ is continuous
at~$q=0$.

The approximate (signed) value of~$|\nabla \cdot {\bf v}|$, denoted
$(\mbox{div } v)$, is given by the fractional change in the volume of
a parcel as it rises a distance~$l$, $[(\rho^\prime)^{-1} -
\rho_0^{-1}]/(\rho_0^{-1})$, divided by the time for the parcel to
rise, $|l/u|$:
\begin{equation}
(\mbox{div } v) = \; -\; \frac{ \mbox{sgn}(l) \, u}{\gamma p +
\gamma_{\rm cr} p_{\rm cr}} \frac{dp_{\rm tot}}{dr},
\label{eq:valdivv} 
\end{equation} 
where~$\mbox{sgn}(l) = l/|l|$.

From equation~(\ref{eq:ene1}), the steady-state thermal-plasma energy equation is
\begin{equation}
\frac{1}{\gamma -1}
 \nabla \cdot({\bf v} p)
 = - p \nabla \cdot {\bf v} + H_{\rm diss} + H_{\rm tc} - R.
\label{eq:e2} 
\end{equation} 
The steady-state cosmic-ray energy equation, from
equations~(\ref{eq:enecr1}) and (\ref{eq:defHcr}) is 
\begin{equation}
\frac{1}{\gamma_{\rm cr} -1}
\nabla \cdot({\bf v} p_{\rm cr})
 = - p_{\rm cr} \nabla \cdot {\bf v} +\nabla \cdot (D_{\rm cr}\nabla p_{\rm cr}) + S.
\label{eq:ecr1} 
\end{equation} 
To obtain averaged equations,  each fluid quantity is written as an average
value plus a turbulent fluctuation:
\begin{equation}
{\bf v}  = \langle {\bf v}  \rangle + \delta {\bf v} ,
\label{eq:avv} 
\end{equation} 
\begin{equation}
p = \langle p \rangle + \delta p, 
\label{eq:avp} 
\end{equation} 
etc. As mentioned above, $\langle {\bf v} \rangle$ is
set equal to zero.
It is assumed that averaged quantities depend only
on the radial coordinate~$r$.
The average $\langle \delta {\bf v} \: \delta p\rangle$,
which is~$\gamma-1$ times the thermal-plasma 
internal-energy flux, is estimated
to be $c_{\rm mix} [u \:\mbox{sgn}(l)\hat{r}] \,\Delta p $,
where~$c_{\rm mix}$ is a
constant of order unity, giving
\begin{equation} 
\langle \delta {\bf v} \: \delta p\rangle =  \hat{r} D Q,
\label{eq:valdpdv} 
\end{equation} 
where
\begin{equation}
D = c_{\rm mix} u |l|
\label{eq:defD} 
\end{equation} 
is the eddy diffusivity, and
\begin{equation}
Q = \frac{\Delta p}{l} = \frac{\gamma p}{\gamma p + \gamma_{\rm cr} p_{\rm cr}}
\frac{dp_{\rm tot}}{dr} - \frac{dp}{dr}.
\label{eq:defQ} 
\end{equation} 
The average $\langle \delta {\bf v} \: \delta p_{\rm cr}\rangle$ is estimated
to be $c_{\rm mix} [u\: \mbox{sgn}(l)\hat{r}] \,\Delta p_{\rm cr} $,
which yields
\begin{equation}
\langle \delta {\bf v} \: \delta p_{\rm cr}\rangle = 
- \langle \delta {\bf v} \: \delta p\rangle .
\label{eq:valdpcrdv} 
\end{equation} 
The average $\langle \delta p  \nabla \cdot \delta {\bf v}\rangle$ is
taken to be $ c_{\rm mix} \Delta p\: (\mbox{div } v) $,
or
\begin{equation} 
\langle \delta p \nabla \cdot \delta {\bf v}\rangle
= \; - \;\frac{DQ}{\gamma p + \gamma_{\rm cr} p_{\rm cr}}
\frac{dp_{\rm tot}}{dr}.
\label{eq:dpddivv}
\end{equation} 
Similarly, 
 $\langle \delta p_{\rm cr}  \nabla \cdot \delta {\bf v}\rangle$ is
taken to be $ c_{\rm mix} \Delta p_{\rm cr}\: (\mbox{div } v) $,
giving
\begin{equation} 
\langle \delta p_{\rm cr} \nabla \cdot \delta {\bf v}\rangle
= - \langle \delta p \nabla \cdot \delta {\bf v}\rangle.
\label{eq:dpcrdivv}
\end{equation} 
Equations~(\ref{eq:valdpcrdv}) and~(\ref{eq:dpcrdivv}) 
reflect the assumption that the
the total-pressure fluctuation in a displaced fluid
parcel vanishes. The mixing length is set equal to
\begin{equation}
|l| = \alpha r,
\label{eq:defalpha}
\end{equation} 
where $\alpha$ is a constant.
Although~$|l|\ll r$ was previously assumed, it is now assumed that 
$\alpha$ is of order unity, an inconsistency that also characterizes
standard mixing-length theory.

Upon averaging
equations~(\ref{eq:e2}) and~(\ref{eq:ecr1}), discarding
terms proportional to the average velocity, 
and dropping the angle brackets $\langle \dots
\rangle$ around averaged quantities, one obtains
\begin{equation}
\frac{1}{(\gamma -1)r^2}
\frac{d}{dr}\left(r^2 D Q\right)
 = 
\frac{DQ}{\gamma p + \gamma_{\rm cr} p_{\rm cr}} \, \frac{dp_{\rm tot}}{dr} 
 + H_{\rm diss} + H_{\rm tc} - R,
\label{eq:e3} 
\end{equation} 
and
\begin{equation}
 \frac{- 1}{(\gamma_{\rm cr} -1)r^2}
\frac{d}{dr}\left(r^2 D Q\right)
 = 
\frac{-DQ}{\gamma p + \gamma_{\rm cr} p_{\rm cr}} \, \frac{dp_{\rm tot}}{dr} 
+ \frac{1}{r^2} \frac{d}{dr} \left(r^2
D_{\rm cr} \frac{dp_{\rm cr}}{dr}\right)
+ S.
\label{eq:ecr2} 
\end{equation} 
In the averages of~$H_{\rm tc}$ and~$R$ the
density and temperature are simply replaced by their average
values. The average and turbulent velocities
are neglected in the momentum equation, giving
\begin{equation}
\frac{dp_{\rm tot}}{dr} = - \rho \frac{d\Phi}{dr},
\label{eq:mom} 
\end{equation} 
where~$\Phi$ is the gravitational potential, which is assumed to be
dominated by a fixed dark matter distribution.
The average value of~$H_{\rm diss}$ is set equal to
\begin{equation}
H_{\rm diss} = \frac{c_{\rm diss} \rho u^3}{l},
\label{eq:defHd}
\end{equation} 
where $c_{\rm diss}$ is a dimensionless constant of order unity.
Equations~(\ref{eq:e3}) through (\ref{eq:defHd}) describe both
convectively stable and convectively unstable regions, but~$D$ and
$H_{\rm diss}$ are effectively zero in stable regions that are not
very close to marginal stability.

A few words on the relationship between the two-fluid convection model
and standard mixing length theory are in order. The left-hand side of
equation~(\ref{eq:e3}) is reminiscent of the divergence of the
convective heat flux that appears in standard mixing length theory and
discussions of forced turbulent mixing in clusters. However, the
convective internal-energy flux $\langle \delta v_r\delta
p\rangle/(\gamma-1)$ on the left-hand side of equation~(\ref{eq:e3})
is not proportional to the specific-entropy gradient. In fact,
$\langle \delta p \:\delta v_r\rangle$ vanishes as $p_{\rm cr}
\rightarrow 0$, because the total pressure perturbation in a fluid
element is assumed to vanish. However, the approach taken in this
paper would be similar to standard mixing length theory in the
$p_{\rm cr}\rightarrow 0$ limit if the average velocity were
retained. In a convectively unstable region, downward falling fluid
elements are denser on average than upwardly rising fluid elements.
The random velocities are thus associated with an inward mass flux. In
standard mixing length theory, the total mass flux vanishes because a
steady state without mass sources or sinks is assumed (Cox \& Giugli
1968). This requires a non-vanishing outward average radial velocity,
\begin{equation}
\langle v_r\rangle = \;-\;\frac{\langle \delta \rho \:\delta v_r\rangle}{\langle
\rho\rangle}.
\label{eq:avvr} 
\end{equation} 
If we add equation~(\ref{eq:e2}) to the dot product of~${\bf v}$ with
the steady-state momentum equation, employ~$\nabla \cdot (\rho {\bf v} ) = 0$,
and take the limit~$p_{\rm cr} \rightarrow 0$, we obtain the total-energy equation
\begin{equation}
0
= - \nabla \cdot \left(\frac{\gamma p {\bf v} }{\gamma-1}
+ \frac{\rho {\bf v} v^2}{2} + \rho {\bf v} \Phi + {\bf F}_V
\right) + H_{\rm tc} - R,
\label{eq:e5} 
\end{equation} 
where $\gamma p {\bf v} /(\gamma-1)$ is the enthalpy flux,
and ${\bf F}_V$ is the viscous energy flux [see, e.g., Rudiger (1989),
equation~8.40].
If one sets
\begin{equation}
\langle \delta \rho \:\delta v_r\rangle = c_{\rm mix}
\Delta \rho \mbox{ sgn}(l) u,
\label{eq:dvdrho} 
\end{equation}
the enthalpy flux associated with the average radial velocity of 
equation~(\ref{eq:avvr}) is
\begin{equation}
\frac{\gamma \langle p\rangle \langle v_r\rangle}{\gamma -1}
= - D\rho T \frac{ds}{dr},
\label{eq:enthflux} 
\end{equation} 
where the angle brackets around average quantities have been dropped
on the right-hand side, $s = C_V \ln (p/\rho^\gamma)$ is the specific
entropy, and $C_V = \epsilon/T$ is the specific heat at constant
volume.  The enthalpy flux in equation~(\ref{eq:enthflux}) is the same
as the heat flux in standard mixing length theory.

Although $\langle v_r\rangle$ must be included to recover standard
mixing length theory in the $p_{\rm cr} \rightarrow 0$ limit, $\langle
v_r \rangle$~can be safely neglected in the energy equation when the
model is applied to clusters. In the limit of vanishing~$\dot{M}$, the
$\langle v_r\rangle \langle p\rangle$~terms in the average of
equation~(\ref{eq:e2}) are smaller than the $\langle \delta v_r \delta
p\rangle$~terms, which can be seen as follows. Using
equations~(\ref{eq:avvr}) and (\ref{eq:dvdrho}), one can write
\begin{equation}
\frac{\langle v_r\rangle \langle p\rangle}{\langle \delta v_r \delta p\rangle}
= \frac{\Delta \rho}{\rho}\left(\frac{\Delta p}{p}\right)^{-1}.
\label{eq:smvr} 
\end{equation} 
This ratio is small for the following reasons.
Convective regions must be close to marginal convective stability to
produce the subsonic turbulent velocities required to balance cooling,
which implies $|\Delta \rho| \ll |l\rho/r|$.  This condition, coupled
with equations~(\ref{eq:valdrho}) and (\ref{eq:valdp}), gives
\begin{equation} \Delta p \simeq \frac{-lp}{C_V} \frac{ds}{dr}.
\label{eq:valdp2} 
\end{equation} 
Equation~(\ref{eq:valdp2}) implies that $|\Delta p| \sim lp/r$, since
$ds/dr$ is observed to be~$\sim C_V/r$ in clusters.  Thus, $|\Delta
\rho/\rho| \cdot |\Delta p/p|^{-1} \ll 1$.\footnote{If one takes~$\rho
g \sim p/r$ and $\nabla \cdot \langle \delta {\bf v} \: \delta
p\rangle \sim \langle \delta v_r \: \delta p\rangle/r$, then $H_{\rm
diss}/\nabla \cdot \langle \delta {\bf v} \: \delta p\rangle \sim
(\Delta \rho\:/\rho)(\Delta p\:/p)^{-1}$, which is again small because
convective regions are near marginal convective stability. Thus,
dissipation of turbulent motions is less important than turbulent
mixing in the model of this paper, as is borne out by numerical
solutions to equations~(\ref{eq:e3}) through (\ref{eq:mom}).  On the
other hand, if an external source of turbulent motions is invoked as
in, e.g., El-Zant et~al~2004, then viscous dissipation may be more
important in comparison to turbulent mixing than in the present
model.}  It is true that if~$\dot{M}$ is sufficiently large, then
the~$\langle v_r\rangle$ terms in the energy equation are comparable
to the radiative-cooling term, as in the cooling flow model. However,
in the model solutions to be presented below,~$\dot{M}$ is a
factor~$10^2-10^3$ smaller than in the cooling-flow
model, so that the neglect of~$\langle v_r\rangle$ is
okay.~\footnote{Very close to cluster center, $\langle v_{\rm
r}\rangle$ becomes large even for small~$\dot{M}$, and the model 
is inaccurate.}

Equation~(\ref{eq:defQ}) can be rewritten
\begin{equation}
Q = - p_{\rm tot} \frac{d\chi}{dr} + \left[\frac{(\gamma-\gamma_{\rm
cr})\chi p_{\rm cr}}{\gamma p + \gamma_{\rm cr}p_{\rm cr}}\right]
\frac{dp_{\rm tot}}{dr},
\label{eq:Q2} 
\end{equation} 
where
\begin{equation}
\chi = \frac{p}{p_{\rm tot}}.
\label{eq:defchi} 
\end{equation} 
One expects~$d\chi/dr > 0$ throughout most of a cluster since cosmic rays contribute
a larger fraction of the pressure nearer the central radio source.
Since~$\gamma\geq \gamma_{\rm cr}$, one expects $Q$ to be negative in clusters.
Thus, if a fluid parcel is displaced inwards ($l<0$), the thermal pressure
inside the displaced parcel is larger than in the surrounding medium ($lQ>0$), 
essentially because the total pressure is the same inside and outside
the parcel and $\chi$ is larger inside. Since the inwardly displaced
parcel gets compressed, the first term on the right-hand side of
equation~(\ref{eq:e3}) ($-\langle \delta p \nabla \cdot \delta
{\bf v} \rangle$) is positive, and acts as an energy source for the
thermal plasma. This source term can be interpreted as in part due to
$pdV$ work on the thermal plasma by the cosmic-ray pressure, which
helps compress the displaced parcel.

Equations~(\ref{eq:e3}) through (\ref{eq:mom}) form a system of two
second-order equations and one first-order equation for~$\rho$, $T$,
and~$p_{\rm cr}$.  Five boundary conditions are required to specify a
solution. Two boundary conditions are obtained by imposing a
density~$\rho_{\rm outer}$ and temperature~$T_{\rm outer}$ at
radius~$r_{\rm outer}$. Two additional boundary conditions are
obtained by taking $dT/dr$ and~$d p_{\rm cr}/dr$ to vanish at the
origin. The fifth condition is obtained by assuming that~$p_{\rm cr}
\rightarrow 0$ as~$r \rightarrow \infty$. This condition is translated
into an approximate condition on~$p_{\rm cr}$ at~$r_{\rm outer}$ as
follows. The value of $r_{\rm outer}$ is chosen to be much greater
than~$r_s$ and $D_0/v_d$, so that for $r\gtrsim r_{\rm outer}$,
$S\simeq 0$ and $D_{\rm cr} \sim v_d r$.  Equation~(\ref{eq:ecr2})
then implies that~$p_{\rm cr} \simeq c_1 + c_2 r^{-2} $ for~$r\gtrsim
r_{\rm outer}$ assuming the ICM is convectively stable and far from
marginal stability. Since~$p_{\rm cr} \rightarrow 0$ as~$r\rightarrow
\infty$, $c_1=0$.  The fifth boundary condition is then taken to be
$dp_{\rm cr}/dr = -2 p_{\rm cr}/r$ at~$r_{\rm outer}$.  Numerical
solutions are obtained using a shooting method. Values are guessed
for~$\rho$,~$T$, and~$p_{\rm cr}$ at~$r=0$ and the equations are
integrated from~0 to~$r_{\rm outer}$.  The guesses are then updated
using Newton's method until the three boundary conditions at~$r_{\rm
outer}$ are met.

\begin{figure}[t]
\vspace{12cm}
\includegraphics{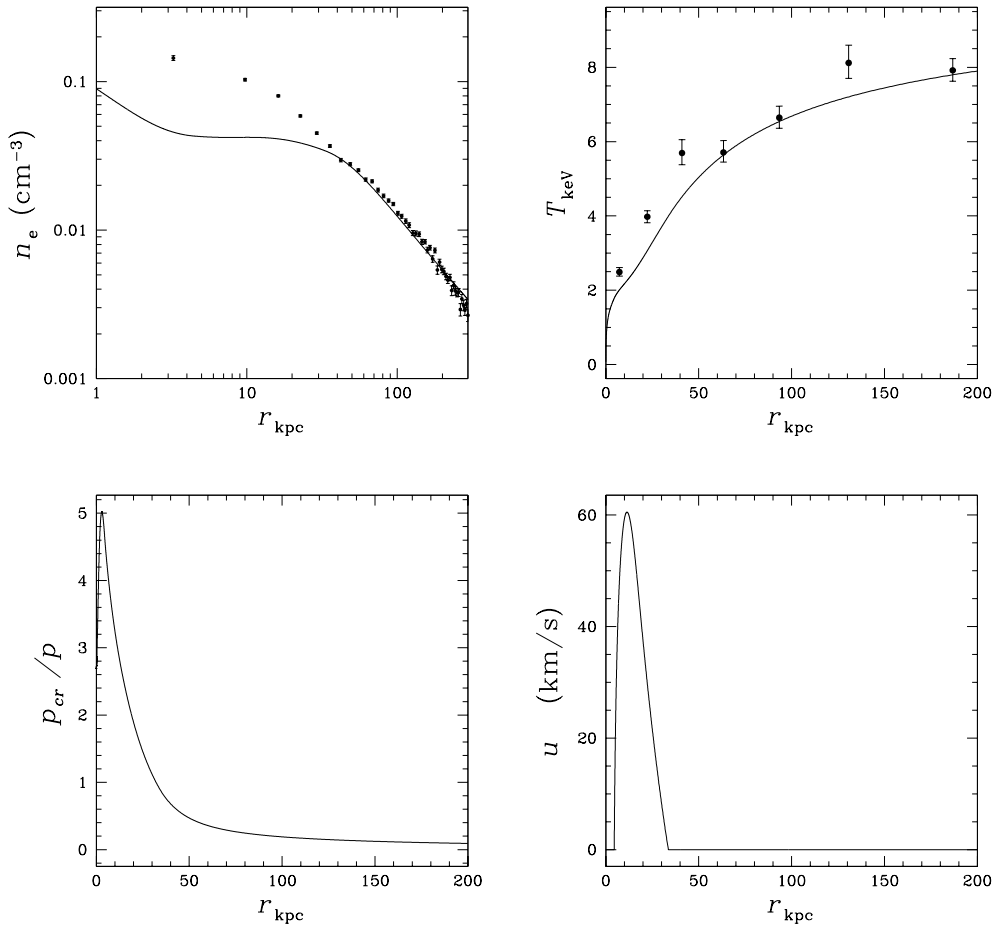}
\caption{\footnotesize Top panels: comparison of model density and
temperature with observations of Abell~478.  Data points and one-sigma
error bars (provided by S.~Allen---see Sun et~al~2003) assume
$\Omega_M = 0.3$, $\Omega_\Lambda =0.7$, and~$H_0 = 70 \mbox{
km}\;\mbox{s}^{-1} \mbox{Mpc}^{-1}$.  Bottom panels give the ratio of
cosmic-ray pressure to thermal pressure and the turbulent velocity as a
function of distance from cluster center.
\label{fig:f1}}
\end{figure}

Figure~\ref{fig:f1} shows an example solution for the following parameters:
$M_{\rm bh} = 10^9 M_\sun$, $\eta = 0.003$, $\gamma = 5/3$,
$\gamma_{\rm cr} = 4/3$, $\alpha =1$, $u_0^\prime = 0.1 \mbox{ km}
\mbox{ s}^{-1} \mbox{kpc}^{-1}$, $r_s = 25$~kpc, $v_d = 18$~km/s,
$D_0 = 10^{28} \mbox{ cm}^2/\mbox{s}$, and~$\theta =1/3$.  The
values $c_{\rm diss} = 0.42$ and~$c_{\rm mix}=0.11$ are adopted based
on a number of previous studies, as discussed by Dennis \&
Chandran~(2004).  The gravitational potential is taken to be
\begin{equation}
\Phi = \frac{v_c^2}{2}\,\ln\left[1 +
\left(\frac{r}{r_c}\right)^2\right]
+  \frac{v_g^2}{2}\,\ln\left[1 +
\left(\frac{r}{r_g}\right)^2\right]
\label{eq:phi} 
\end{equation} 
with $v_c = 1150$~km/s, $r_c= 40$~kpc, $v_g = 400$~km/s,
and~$r_g = 1$~kpc. The term proportional to~$v_c^2$
($v_g^2$) represents the cluster
(central-galaxy)~potential. A mean molecular weight per
electron~$\rho/n_e m_H$ of~1.18 is assumed, where~$m_H$ is
the mass of a hydrogen atom, and the ratio~$n_i/n_e$ is set
equal to~0.9.\footnote{These values are appropriate for a
hydrogen fraction $X=0.7$ and a helium fraction $Y=0.28$,
with fully ionized hydrogen and helium (Zakamska \& Narayan
2003).}  The values $\rho_{\rm outer} = 5.38\times 10^{-3}
\mbox{ cm}^{-3}$ and~$T_{\rm outer} = 7.91$~keV at~$r_{\rm
outer} = 200$~kpc are determined by linear interpolation
between data points for Abell~478.  Density and temperature
data for Abell~478 from Chandra observations, provided by
S.~Allen, are also plotted in figure~\ref{fig:f1}. Abell~478
is a highly relaxed, x-ray luminous cluster, with a
cooling-flow-model mass accretion rate of~$\sim 1000 M_\sun
\mbox{ yr}^{-1}$, and a comparatively weak central radio
source (Allen et~al~1993, Sun et~al~2003).

It can be seen from the figure that the the model density is too
small for~$r\lesssim 50$~kpc. This discrepancy arises because the
specific entropy of the thermal plasma increases outwards, and thus
the cosmic-ray pressure gradient must be large in order for the ICM to
be convectively unstable and for turbulent heating to help balance
radiative cooling. When cosmic rays contribute a large fraction of the
total pressure support, the thermal pressure gradient does
not have to be as large, and thus the central plasma density does not
have to be as large. A variety of model parameters have been
investigated with similar results.

\begin{figure}[t]
\vspace{12cm}
\includegraphics{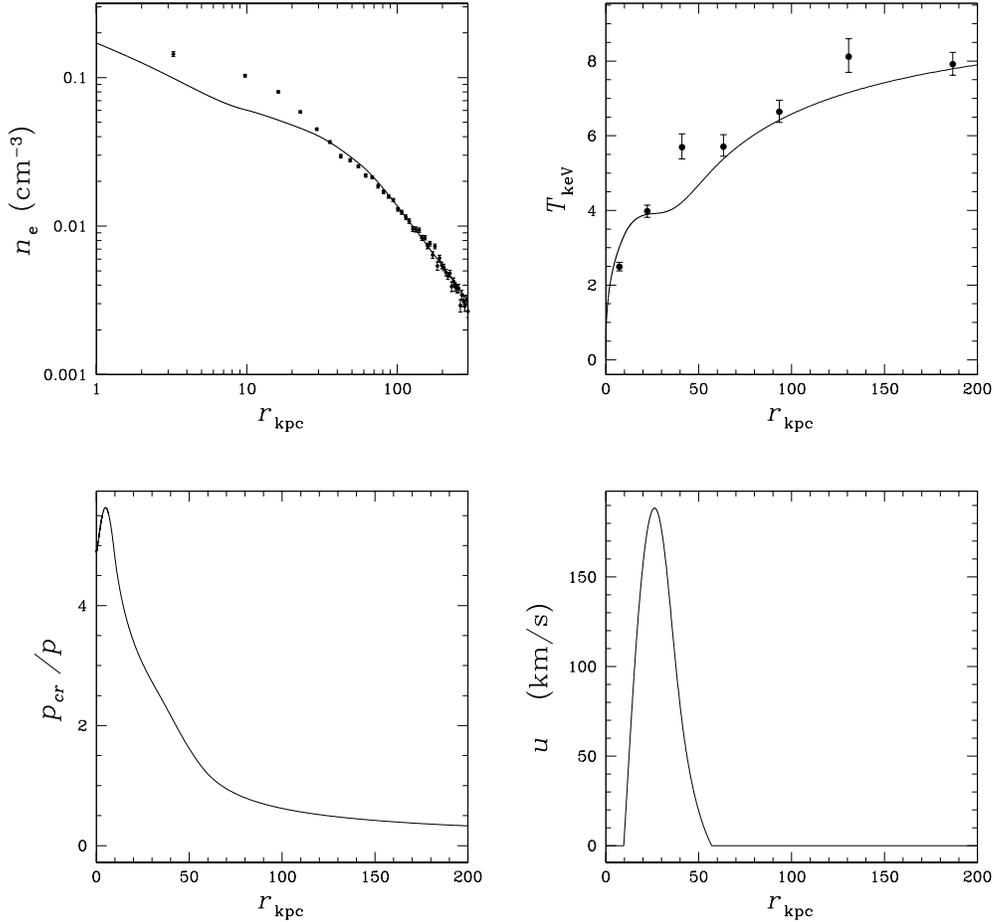}
\caption{\footnotesize Same as figure~\ref{fig:f1}, but with
different model parameters, as described in the text.
\label{fig:f2}}
\end{figure}

The turbulent velocity in figure~\ref{fig:f1} is also surprisingly
small.  This is in part because the model underestimates the plasma
density in the central~$50$~kpc, which weakens radiative cooling and
the need for turbulent heating. Figure~\ref{fig:f2} shows a numerical
solution for the parameters $v_c = 1500$~km/s, $x_c = 10$~kpc, $v_g =
0$, $\theta = 0.37$, $v_d = 20$~km/s, and $D_0 = 3\times 10^{28}
\mbox{ cm}^2/\mbox{s}$, with other parameters the same as in
figure~\ref{fig:f1}.  The enhanced gravitational acceleration
increases the density and radiative cooling, which results in the need
for greater turbulent heating and a larger~$u(r)$. These parameters,
however, are unrealistic since they imply a 
velocity dispersion of~$\sim 1000$~km/s within the central galaxy
at~$r=10$~kpc, which is much larger than observed velocity dispersions
in giant elliptical galaxies.

A further difficulty with the model is that a critical point is
encountered for sufficiently large~$v_d$ and/or~$D_0$, or for
sufficiently small~$\theta$. It can be shown that a sufficient
condition for avoiding a critical point is that $(\gamma_{\rm
cr}-1)D_{\rm cr}/(\gamma-1)$ is everywhere less than $\kappa_T/n_e
k_B$, the diffusion coefficient of heat-carrying electrons, and the
models plotted in figures~\ref{fig:f1} and~\ref{fig:f2} satisfy this
inequality.  The potential for a critical point raises a number of
interesting questions, but these lie beyond the scope of this paper.

For reference, in the numerical solution plotted in figure~\ref{fig:f1},
the Bondi accretion rate is $1.2 M_\sun \mbox{ yr}^{-1}$, $L_{\rm cr} = 2.1 \times
10^{44}$~ergs/s, the radiative luminosity out to 200~kpc is~$1.9
\times 10^{45}$~erg/s, and the radiative luminosity out to 600~kpc
is $4.9\times 10^{45}$~ergs/s.
For the model plotted in figure~\ref{fig:f2},
the Bondi accretion rate is~$5.0 M_\sun \mbox{ yr}^{-1}$,
$L_{\rm cr} = 8.5 \times 10^{44}$~erg/s, 
the radiative luminosity out to 200~kpc is
$2.2 \times 10^{45}$~erg/s, and the radiative
luminosity out to 600~kpc is $4.5\times 10^{45}$~erg/s.
(The luminosity out to 600~kpc is smaller in the
model of figure~\ref{fig:f2} because the larger gravitational
acceleration causes the density to drop off faster at large~$r$.)

\section{Summary}
\label{sec:dis} 

AGN feedback is a promising explanation for the heating of intracluster
plasmas on both observational and theoretical grounds: there is an
active radio source at the center of almost every strongly cooling
cluster (Eilek 2003), and AGN feedback in combination with thermal
conduction can lead to globally stable equilibria (Rosner \& Tucker
1989, Ruszkowski \& Begelman 2002). The recent detection of moderately
subsonic plasma motions in the Perseus cluster (Churazov
et~al~2004) further suggests that
turbulent mixing and/or turbulent dissipation plays an important role in
heating the intracluster medium. This observation also poses a
theoretical challenge. For velocities in the range of those observed,
turbulent heating can either overwhelm radiative cooling or be too
small to offset cooling, depending on the precise value of the rms
velocity~$u$ and the velocity length scale.  If turbulent heating
indeed balances radiative cooling, an explanation is needed for
how~$u$ is  fine-tuned to the required value.

The model presented in this paper connects AGN feedback to turbulent
heating, and seeks to explain the fine-tuning of~$u$. In the model,
radiative cooling of intracluster plasma is balanced by a combination
of turbulent heating and thermal conduction. The turbulence is
generated by convection, which in turn is produced by the buoyancy of
cosmic rays generated by a central radio source.  Convection is a
natural way for a central relativistic jet to generate turbulence due
to the jet's small momentum-flux-to-energy-flux ratio: rather than a
relativistic outflow stirring intracluster plasma like an oar in
water, the radio source inflates the central region, allowing gravity
to generate plasma momentum after parcels of plasma and cosmic rays in
the central region become buoyant. A two-fluid (plasma and cosmic ray)
mixing length theory is developed to treat convection in the~ICM. By
linking the turbulence amplitude to the mass accretion rate and AGN
feedback luminosity, the model provides an explanation for how the
turbulent velocity achieves the value required for heating to balance
cooling. Although stability is not investigated in this paper, it is
expected that the model equilibria are stable for the same reasons as
other AGN-feedback/thermal-conduction models.

The model provides equilibrium solutions for $n_e(r)$, $T(r)$, $p_{\rm
cr}(r)$, and~$u(r)$ once the plasma temperature and density are
specified at some suitably large outer radius. For typical cluster
parameters, the ICM stays fairly close to marginal convective
stability in convective regions, implying subsonic turbulent
velocities. The principal shortcoming of the model is that the model
density is too small within the central~$\sim 50$~kpc, where the model
ICM is convective. The reason is that the cosmic-ray pressure gradient
must be large for cosmic-ray buoyancy to generate convection. This in
turn reduces the thermal pressure gradient in the central region and
decreases the central plasma density.  A more realistic model of
convection in the ICM in which relativistic and thermal plasmas are
only partially mixed may have greater success.

\acknowledgements
I thank Steve Allen for providing the data for Abell~478, 
Eric Blackman, Eliot Quataert, and the anonymous referee
for providing helpful comments on an 
earlier version of this paper, and 
Steve Allen, Steve Cowley, Ue-li Pen, and
Steve Spangler for helpful discussions.
This work was supported by NSF grant AST-0098086 and DOE grants
DE-FG02-01ER54658 and DE-FC02-01ER54651 at the University of Iowa.

\references

Allen, S. W., Fabian, A. C., Johnstone, R. M., White, D. A., Daines, S. J.,
Edge, A. C., \& Stewart, G. C. 1993, MNRAS, 262, 901

Begelman, M. C. 2001, in ASP Conf. Proc., 240, {\em Gas and Galaxy Evolution},
ed. J. E. Hibbard, M. P. Rupen, \& J. H. van Gorkom
(San Fransisco: ASP), 363

Begelman, M. C. 2002, in ASP Conf. Proc., 250, {\em Particles and Fields
in Radio Galaxies}, ed. R. A. Laing, \& K. M. Blundell 
(San Fransisco: ASP), 443

Bessho, N., Maron, J., \& Chandran, B. 2004, in preparation

Binney, J., \& Tabor, G. 1995, MNRAS, 276, 663

Birzan, L., Rafferty, D., McNamara, B., Wise, M., \& Nulsen, P. 2004,
ApJ, 607, 800

B\"{o}hringer, H. et~al 2001, A\&A, 365, L181

B\"{o}hringer, H., Matsushita, K., Churazov, E., \& Finoguenov, A. 2003,
in {\em The Riddle of Cooling Flows and Clusters
of Galaxies}, ed. Reiprich, T., Kempner, J., \& Soker, N., E3,
{\bf http://www.astro.virginia.edu/coolflow/proc.php}

B\"{o}hringer, H., \& Morfill, G. 1988, ApJ, 330, 609

Bondi, H. 1952, MNRAS, 112, 159

Brighenti, F., \& Mathews, W. 2002, ApJ, 573, 542

Br\"{u}ggen, M., Kaiser, C., Churazov, E., \& Ensslin, T. 2002, MNRAS, 331, 545

Chandran, B., \& Cowley, S. 1998, Phys. Rev. Lett., 80, 3077

Chandran, B., Maron, J. 2004, ApJ, 602, 170

Cho, J., Lazarian, A., Honein, A., Knaepen, B., Kassinos, S., \& Moin, S. 2003,
ApJ, 589, L77

Churazov, E., Br\"{u}ggen, M., Kaiser, C., B\"{o}hringer, H., \& Forman, W. 2001,
ApJ, 554, 261

Churazov, E., Forman, W., Jones, C., \& B\"ohringer, H. 2000, A\&A, 356, 788

Churazov, E., Forman, W., Jones, C., Sunyaev, R., \& B\"{o}hringer, H. 2004,
MNRAS, 347, 29

Churazov, E., Sunyaev, R., Forman, W., \& B\"{o}hringer, H. 2002, MNRAS, 332, 729

Ciotti, L.,  \& Ostriker, J. 1997, ApJ, 487, L105

Ciotti, L., \& Ostriker, J. 2001, ApJ, 551, 131

Cox, J., \& Guili, R. 1968, {\em Principles of Stellar Structure}
(New York: Gordon and Breach)

Dennis, T., \& Chandran, B. 2004, ApJ, submitted

Eilek, J. 2003, in {\em The Riddle of Cooling Flows and Clusters
of Galaxies}, ed. Reiprich, T., Kempner, J., \& Soker, N., E13,
{\bf http://www.astro.virginia.edu/coolflow/proc.php}

El-Zant, A., Kim, W., \& Kamionkowski, M. 2004, astro-ph/0403696

Ensslin, T. 2003, A\&A, 399, 409

Fabian, A. C. 1994, Ann. Rev. Astr. Astrophys., 32, 277

Fabian, A., Sanders, J., Allen, S., Crawford, C., Iwasawa, K., Johnstone, M.,
Schmidt, R., \& Taylor, G. 2003, MNRAS, 344, L43

Fabian, A., Voigt, L., \& Morris, R. 2002, MNRAS, 335, L71

Kim, W., \& Narayan, R. 2003a, ApJ, 596, 889

Kim, W., \& Narayan, R. 2003b, ApJ, 596, L13

Kronberg, P. 1994, Rep. Prog. Phys., 57, 325

Loewenstein, M., \& Fabian, A. 1990, MNRAS, 242 120

Loewenstein, M., Zweibel, E., \& Begelman, M. 1991, ApJ, 377, 392

Maron, J., Chandran, B., \& Blackman, E. 2004, PRL, accepted

Mathews, W., Brighenti, F., Buote, D., \& Lewis, A. 2003, 596, 159

Molendi, S., \& Pizzolatao, F. 2001, ApJ, 560, 194

Narayan, R., \& Kim, W. 2004, astro-ph/0402206

Narayan, R., \& Medvedev, M. 2001, ApJ, 562, 129

Nulsen, P. 2003,
in {\em The Riddle of Cooling Flows and Clusters
of Galaxies}, ed. Reiprich, T., Kempner, J., \& Soker, N., E30,
{\bf http://www.astro.virginia.edu/coolflow/proc.php}

Owen, F., \& Ledlow, M. 1997, ApJS, 108, 410

Pen, U., Matzner, C., \& Wong, S. 2003, ApJ, 596, L207

Peterson, J. R., et al 2001, A\&A, 365, L104

Peterson, J. R., Kahn, S., Paerels, F., Kaastra, J., Tamura, T., Bleeker, J.,
Ferrigo, C., \& Jernigan, J. 2003, ApJ, 590, 207

Quataert, E., \& Narayan, R. 2000, ApJ, 528, 236

Reynolds, C. S. 2002, in ASP Conf. Proc., 250, {\em Particles and Fields
in Radio Galaxies}, ed. R. A. Laing, \& K. M. Blundell 
(San Fransisco: ASP), 449

Reynolds, C. S., McKernan, B., Fabian, A., Stone, J., \& Vernaleo, J. 2004,
MNRAS, submitted

Rosner, R., \& Tucker, W. 1989, ApJ, 338 761

Rudiger, G., 1989, {\em Differential rotation and stellar convection}
(Berlin: New York)

Ruszkowski, M., \& Begelman, M. 2002, 581, 223

Soker, N. 2003, MNRAS, 342, 463

Spitzer, L. 1962, Physics of Fully Ionized Gases (2d ed.; New York: Wiley)

Sun, M., Jones, C., Murray, S., Allen, S., Fabian, A., \& Edge, A. 2003,
ApJ, 587, 619

Tabor, G., \& Binney, J. 1993, MNRAS, 263, 323

Tamura, T. et~al 2001, A\&A, 365, L87

Taylor, G., Fabian, A., Allen, S. 2002, MNRAS, 334, 769  

Taylor, G., Govoni, F., Allen, S., Fabian, A. 2001, MNRAS, 326, 2

Tozzi, P., \& Norman, C. 2001, ApJ, 546, 63

Voigt, L. M., Schmidt, R. W., Fabian, A. C., Allen, S. W., \& Johnstone, R. M. 2002,
{\em Mon. Not. R. Astr. Soc.}, 335, L7

Voigt, L., \& Fabian, A. 2004, MNRAS, 347, 1130

Zakamska, N., \& Narayan, R. 2003, ApJ, 582, 162

\end{document}